\documentclass{article}
\usepackage{cite,bm,graphicx,hyperref}
\begin{document}

\baselineskip=12pt
\def\p{\partial}
\def\be{\begin{equation}}
\def\bea{\begin{eqnarray}}
\def\ee{\end{equation}}
\def\eea{\end{eqnarray}}
\def\bearst{\begin{eqnarray*}}
\def\eearst{\end{eqnarray*}}
\def\dbar{\bar \partial}
\def\nn{\nonumber}
\def\ll{\lambda}
\def\l{\label}
\def\D{\Delta}
\def\o{\over}
\def\E{{\rm e}}
\def\peleven{\parbox{14cm}}
\def\peffec{\peight{\bearst\eearst}\hfill\peleven}
\def\pspace{\peight{\bearst\eearst}\hfill}
\def\ptwelve{\parbox{15cm}}
\def\peight{\parbox{8mm}}
        \def\sl{\hbox{{\rm sl}}}
     \def\half{{1 \over 2}}
     \def\Uq#1{{\rm U}_q \left( #1 \right) }
     \def\Um#1{{\rm U}_{\mu} \left( #1 \right) }
     \def\U#1{{\rm U} \left( #1 \right) }
     \def\D{{\Delta }}
     \def\P{{\Phi}}
     \def\p{{\phi}}
     \def\si{{\psi}}
     \def\d{{\partial}}
     \def\R{{\bf R}}
     \def\s{\sum}
     \def\pr{\prod}
     \def\a{\alpha}
     \def\eps{\epsilon}
     \def\HG{\H_{"G"}}
     \def\Ts{\tilde{\psi}}
     \def\Si{\Psi}
     \def\Raw{\Rightarrow}
     \def\raw{\rightarrow}
     \def\da{\dag}
     \def\lm{\lambda}
     \def\sq{\sqrt}
     \def\D{\Delta}
     \def\b{\beta}
     \def\i{\imath}
     \def\im{\rm {\it Im}}
     \def\re{\rm {\it Re}}
     \def\ah{\hat{A}}
     \def\bh{\hat{B}}
     \def\ch{\hat{C}}
     \def\eh{\hat{D}}
     \def\c{\cdot}
     \def\cs{\cdots}
     \def\e{\epsilon}
     \def\g{\gamma}
\def\CN{{\cal N}}
\def\CM{{\cal M}}
\def\CA{{\cal A}}
\def\CB{{\cal B}}
\def\CD{{\cal D}}
\def\CO{{\cal O}}
\def\CD{{\cal D}}
\def\W{{\cal W}}
\def\lk{\left [}
\def\rk{\right ]}
\def\kt#1{\mid{{#1}}>}
\def\br#1{<{#1}\mid}
\def\z#1{z_{#1}}
\def\de#1{\Delta_{#1}}
\def\ds{\Delta_\Psi}
\def\df{\Delta_\Phi}
\def\da{\Delta_A}
\def\db{\Delta_B}
\def\dc{\Delta_p}
\def\I{\rm {I\kern-.3em I}}
\def\C{\rm {I\kern-.520em C}}
\def\R{\rm {I\kern-.3em R}}
\def\CZ{\rm {Z\kern-.4em Z}}
\def\str{$sl(2,\C)$}
 \def\ps#1{\psi({#1})}
\def\unit{\rm {1\kern-.4em 1}}
\def\f{\frac}
    \def\ff{\f{1}{2}}
    \def\pp{<\Psi(\z1)\Psi(\z2)>}
\def\abf{<A(z_1)B(z_2)\Phi(z_3)>}
\def\abs{<A(z_1)B(z_2)\Psi(z_3)>}
\def\ab#1{\vert #1 \vert}
\def\kp{\kt{\de{p}}}
\def\kpp{\kt{{\D'}_p}}
\def\kk{{\bf k}}
\def\E{{\rm e}}
\def\peleven{\parbox{11cm}}
\def\peffec{\peight{\bearst\eearst}\hfill\peleven}
\def\pspace{\peight{\bearst\eearst}\hfill}
\def\ptwelve{\parbox{12cm}}
\def\peight{\parbox{8mm}}

\title{ Zero Tension Kardar-Parisi-Zhang Equation in
$(d+1)$--Dimensions }
    \author{ A. Bahraminasab $^a$, S.M.A. Tabei $^a$,
    A. A. Masoudi $^c$, \\ \\ F. Shahbazi $^{d}$ and M. Reza Rahimi
    Tabar $^{a,b}$ \\ \\
 $^a$  Department of Physics, Sharif University of
Technology, P.O. Box 11365-9161, Tehran, Iran \\
$^b$ CNRS UMR 6529, Observatoire de la C$\hat o$te d'Azur, BP
4229, 06304 Nice Cedex 4, France \\
 $^c$  Department of Physics, Alzahra University, Tehran, 19834,
 Iran\\
$^d$ Department of Physics, Isfahan University of Technology,
Isfahan,Iran }

\maketitle

\begin{abstract}
The joint probability distribution function (PDF) of the height
 and its gradients is derived for a zero tension $d+1$-dimensional
 Kardar-Parisi-Zhang (KPZ) equation.
It is proved that the height`s PDF of zero tension KPZ equation
shows lack of positivity after a finite time $t_{c}$. The
properties of zero tension KPZ equation and its differences with
the case that it possess an infinitesimal surface tension is
discussed. Also potential relation between the time scale $t_{c}$
and the singularity time scale $t_{c, \nu \rightarrow 0}$ of the
KPZ equation with an infinitesimal surface tension is
investigated.
\end{abstract}
\maketitle
\newpage



\section{Introduction}

Studying the morphology, formation and growth of interfaces has
been one of the recent interesting fields of study because of its
high technical and rich theoretical advantages. On account of the
disorder nature embedded in the surface growth, stochastic
differential equations have been used as a suitable tool for
understanding the behavior of various growth processes. Such
equations typically describe the interfaces at large length
scales, which means that the short length scale details has been
neglected in order to derive a continuum equation by focusing on
the coarse grained properties.
 A great deal of recent theoretical modeling has
been started with the work of Edward and Wilkinson [1] describing
the dynamics of height fluctuations by a simple linear stochastic
equation. By adding a new term proportional to the square of the
height gradient, Kardar, Parisi and Zhang made an appropriate
description for lateral interface growth [2].
 The $d+1$-dimensional
forced KPZ equation is written as

\begin{equation}\label{}
\frac{\partial h}{\partial t} - \frac{\alpha}{2}(\nabla h)^2 = \nu
\nabla^2h + f
\end{equation}
where $h(\bf x,t)$ specifics the surface height at point $\bf x$ (
$d$-dimensional vector ) and $\alpha \geq 0$. The force $f$ is a
zero mean, statistically homogeneous, white in time, Gaussian
process which it's covariance would be
\begin{equation}\label{}
\langle f( {\bf x},t)f({\bf x'},t')\rangle=2D_0D({\bf
x-x'})\delta(t-t')
\end{equation}
Typically the spatial correlation of the forcing is considered to
be a delta function, mimicking the short length correlation. Here
the spatial correlation is considered as
\begin{equation}\label{}
D({\bf x-x'})=\frac{1}{({\pi\sigma^2})^{d/2}}\exp(-\frac{(\bf
{x-x'})^2}{\sigma^2})
\end{equation}
where $\sigma$ is  much less than the system size $L$, i.e.
$\sigma << L$, which  represents a short range character for the
random forcing. It is useful to rescale the KPZ equation as
$h'=h/h_0$, ${\bf r'=r/r_0}$ and $t'=t/t_0$. If we let
$h_0=(\frac{D_0}{\nu })^{1/2}$ and $t_0^2= \frac{r_0}\nu $, where
$r_0$ is a characteristic length, all of the parameters can be
eliminated, except the coupling constant $g= \frac{\alpha
^2D_0}{\nu ^3}$. The limit $g\rightarrow \infty $ (or zero tension
limit, $\nu \rightarrow 0$), is known as the strong coupling
limit [13,14]. Although originally, this equation appeared as a
model for surface growth [3], it is mostly used today in polymer
physics[4], Burgers turbulence [5], nonlinear acoustics[6] and
cosmology [7], etc.

 The nonlinearity of
the KPZ equation includes the possibility of singularity
formations in a finite time as a result of the local minima
instability.  Meaning  that there is a competition between the
diffusion smoothing effect ( the Laplacian term), and the
enhancement of non-zero slopes.
 Let us mention the main
properties of the KPZ equation for the cases $\nu \rightarrow 0$
and $\nu = 0$. In the {\it zero tension limit} ( $\nu \rightarrow
0$) the KPZ equation has the following properties: (i) the
unforced KPZ equation develops singularities for given
dimensions. In one spatial dimension the sharp valleys are
developed in a finite time $t_{c, \nu \rightarrow 0}$. The
geometrical picture consists of a collection of sharp valleys
intervening a series of hills in the stationary state [8].
 In two spatial
dimensions the KPZ equation develops three types of singularities
in finite time. The first singularities are sharp valley lines
with finite lengths, which the height gradients are discontinues
while crossing the valley lines.
 The second type are the end points of the
sharp valley lines. As time goes on these sharp valley lines hit
each other and the crossing point of two valley lines produces a
valley node. Generically these end points disappear at large time
scales and only a network of sharp valley lines will survive
[5,8,9]. In three and higher dimensions, the structure of the
singularities can be more complicated. For instant, in three
dimensions the singularities are, in the language of Burgers
equation, shock surfaces, its boundaries,  the intersection line
of two shock surfaces and finally the point which three shock
surfaces meet each other. The height gradients are discontinues
while crossing the shock surfaces.
 A complete classification of the KPZ
singularities, by considering the metamorphosis of singularities
as time elapses, has been done in [10].
 (ii) Similarly, for
white in time and smooth in space forcing, in the zero tension
limit singularities will be developed  in a finite time in any
given spatial dimensions. For instance, in two spatial dimensions
the sharp valley lines are smooth curves where in the stationary
state the sharp valley lines produces a curvilinear hexagonal
lattice [9,15] (see Figs. (1)). In three dimensions it can be
shown that the shape of the polyhedra tiling the space can not be
determined uniquely, nevertheless, the minimal polyhedra in $d=3$
will have $24$ vertices [9].

The KPZ equation with vanishing surface tension  ( $\nu =0$ )
produces multi-valued solution after time scale $t_{c, \nu=0}$
[5].  In Fig.2 we demonstrate the multi-valued solution of the
unforced zero tension KPZ equation in two dimensions. We have
used the Lagrangian method to simulate the KPZ equation with $\nu
=0$ and initial condition $h(x,y,0) = sin(x) sin(y)$ [5]. The
sinusoidal function is a typical initial condition and has been
used only for simplicity.
 With
this initial condition  the time scale that the KPZ equation
produces the multivalued solution can be found exactly as $t_{c,
\nu =0} = 1$ [5]. As shown in Fig.(2-a), it is evident that for
time scales $t < t_{c, \nu =0} $  the height field is
single-valued. At the time scale $t=t_{c, \nu =0}$ the height
field become singular (see Fig.(2-b)). Finally in Fig.(2-c) we
have plotted the height field for time scale $t > t_{c, \nu =0}$
in which the height is multi-valued. The singularities in the
limit $\nu \rightarrow 0$ can be constructed from multi-valued
solutions of the KPZ equation with $\nu =0$ by Maxwell cutting
rule  [5], which makes the discontinuity in the derivative of
height field. Indeed the Maxwell cutting rule states, that for
$\bf x$`s which the height field $h(x,t)$ becomes multi-valued,
the physical solution can be chosen so that, $h_{ph}({\bf x},t) =
Max \{ h_1 ({\bf x},t), h_2 ({\bf x},t), \cdots, h_N ({\bf x},t)
\}$, where $N$ is the number of the multiple solutions of the
field $h({\bf x},t)$ at position $\bf x$.

\begin{figure}[htbp]
  \includegraphics[width=0.5\linewidth]{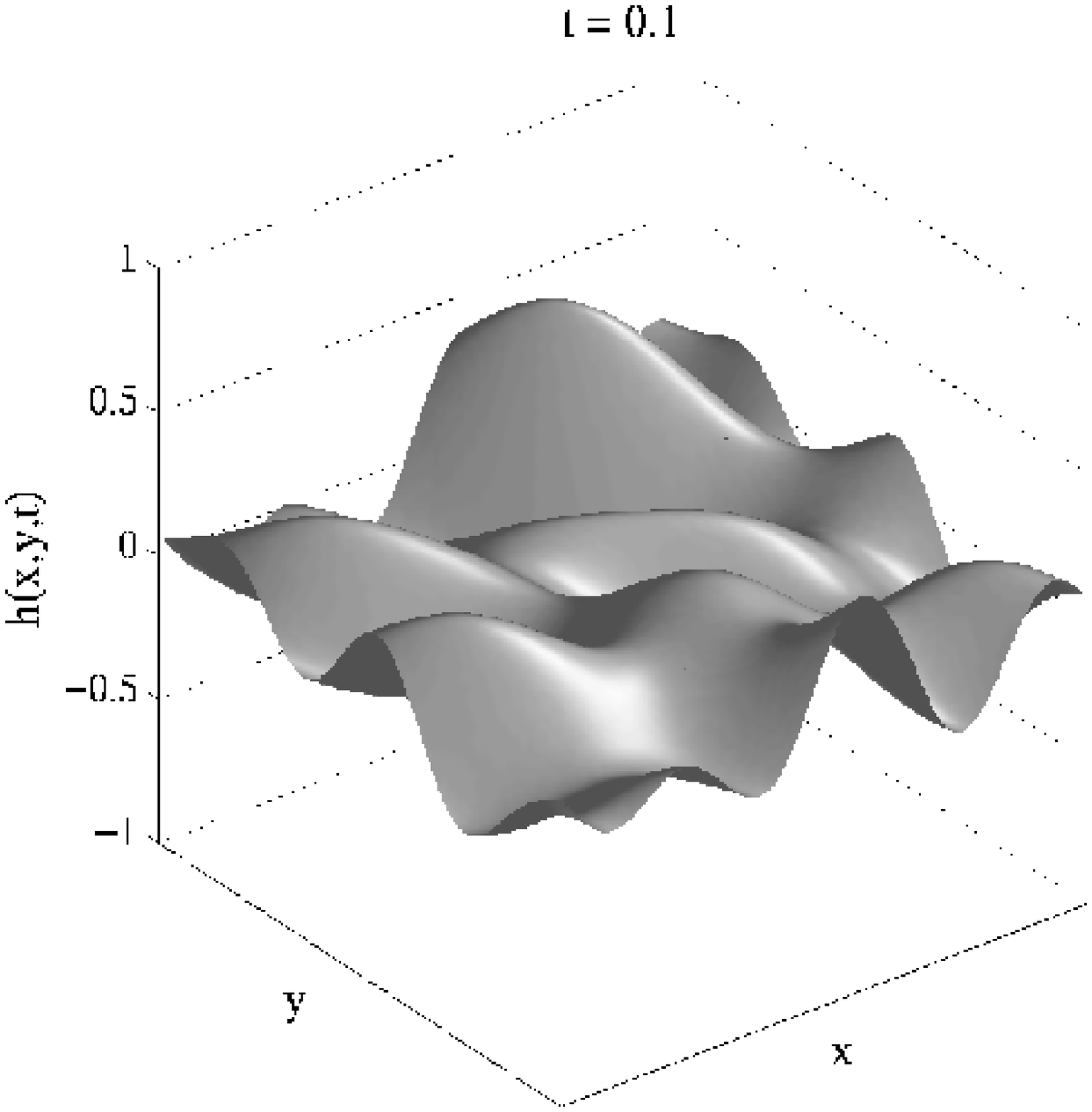}%
  \includegraphics[width=0.5\linewidth]{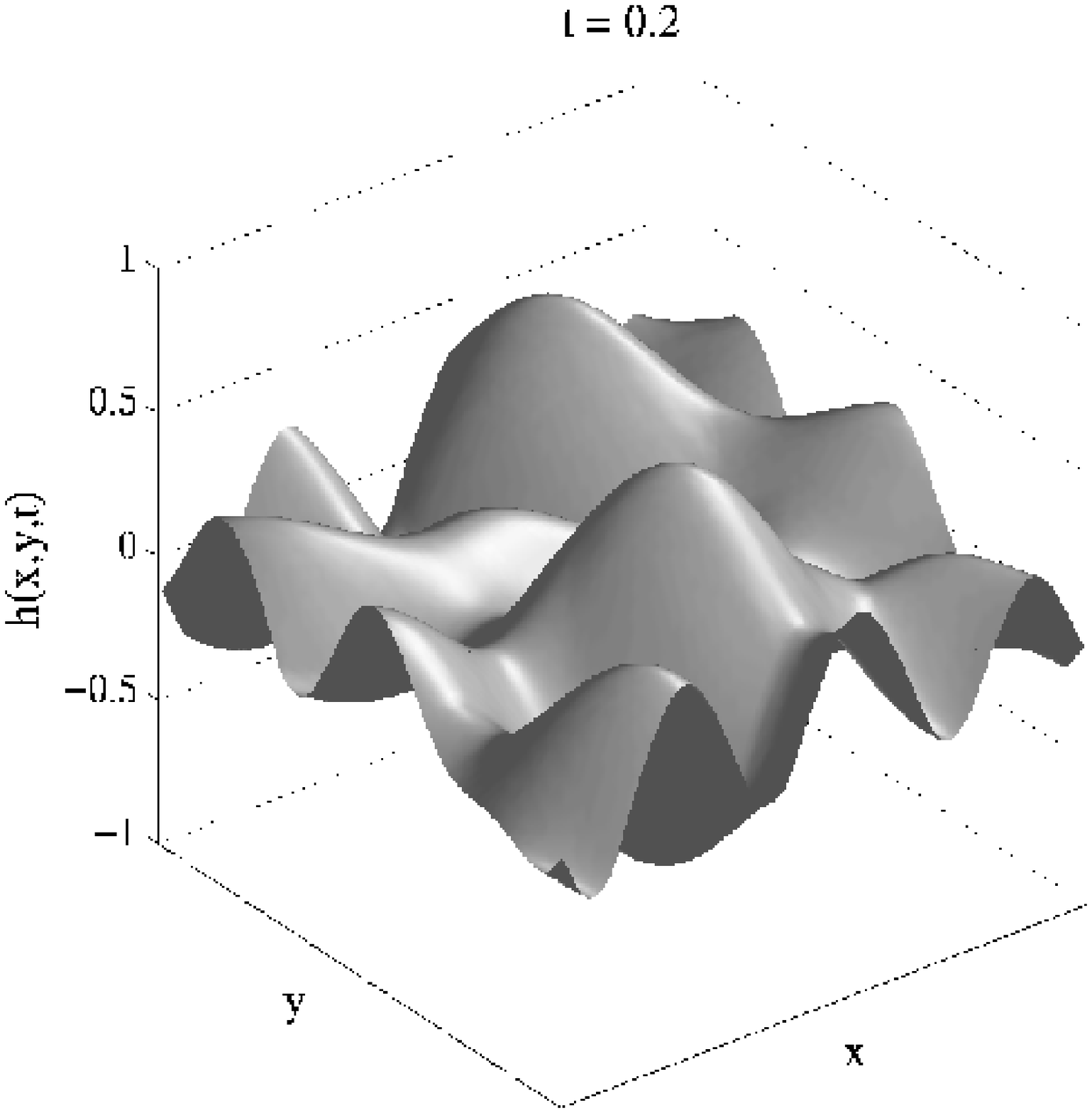}
\includegraphics[width=0.5\linewidth]{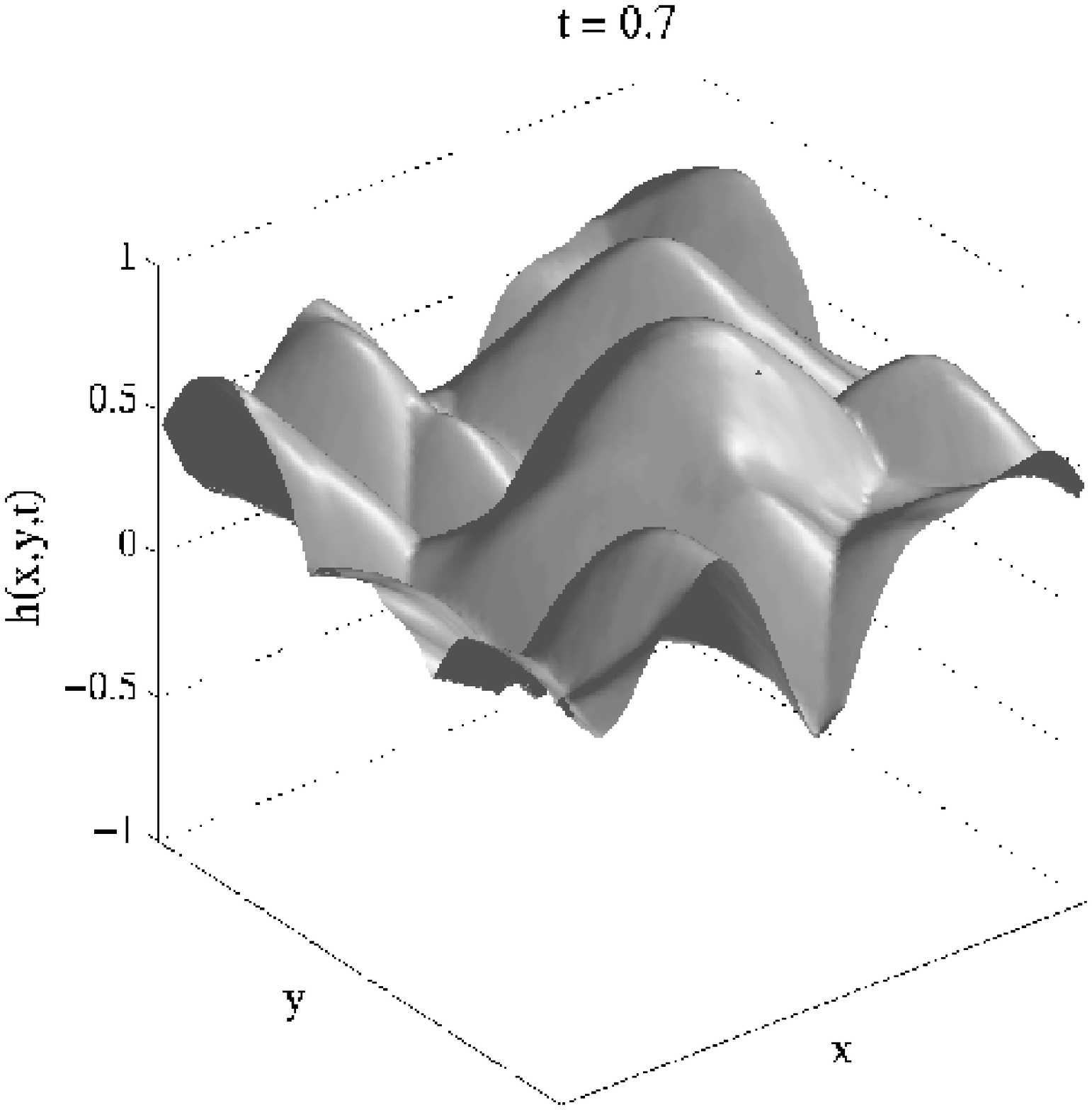}
  \caption[Estimated drift coefficient $D^{(1)}(h_r, r)$]%
  {Different time snapshots of height fields of randomly
driven two dimensional KPZ equation in the zero tension limit (
$\nu \rightarrow 0$). Two upper snapshots are belong to the height
fields before singularity time scales $t_{c , \nu \rightarrow 0}$.
In the lower figure which is for time scales $t
> t_{c, \nu \rightarrow 0}$, the field $h(x,y)$ is not differentiable in singularity curves
[12].  In this simulation the relation between the forcing length
scale $\sigma$ and the sample size $L$, is  $\sigma \simeq L/3$.
The forcing strength $D_0$ is equal to unity. }
  \label{fig:D1}\label{fig:D2}
\end{figure}

In this paper, an exact master equation is derived from the {\it
zero tension} ($\nu =0$) KPZ equation for the joint probability
distribution function (PDF) of height and its gradients.
 The master equation
enables us to determine the time evolution of the PDF of $ h -
\bar h $ and all of the moments $ < (h - \bar h)^n > $. It is
proved that the derived height`s PDF for the $\nu = 0$ case, shows
lack of positivity after a finite time $t_{c}$. Potential relation
between the $t_{c}$ and the singularity time scale ( $t_{c, \nu
\rightarrow 0}$) of the KPZ equation having an infinitesimal
surface tension is discussed. Details of calculations are
presented in the appendices A,B and C.
\\

\section{ Joint PDF of height and its gradients for zero tension KPZ equation   }

  Let us define $P(\widetilde{h},u_i,p_{ij},t)$ as the
joint PDF of $\tilde h = h - \bar h$, $u_i= h_{x_i}$ and
$p_{ij}=h_{x_i x_j}$ where $i,j= 1,2, \cdots,d$. Using the zero
tension KPZ equation, it is shown in appendix A, that the
$P(\widetilde{h},u_i,p_{ij},t)$ satisfies the following equation

\begin{eqnarray}
P_t=\gamma(t)P_{\widetilde{h}}+\frac{\alpha}{2}\sum_l
u_l^2P_{\widetilde{h}}-\alpha (d+2)\sum_lp_{ll}P\hspace{.8cm}\nonumber\\
-\alpha\sum_{l,k\leq m}p_{lk}p_{lm}P_{p_{km}}+
k(0)P_{\widetilde{h}\widetilde{h}}
-k''(0)\sum_{l}P_{u_lu_l}\hspace{.2cm}\nonumber\\
+2k''(0)\sum_lP_{\widetilde{h}p_{ll}} +k''''(0)\sum_{l\leq
k}P_{p_{lk}p_{lk}}\nonumber\\
-2k''''(0)\sum_{l<k }P_{p_{ll}p_{kk}}.
\end{eqnarray}

where $\gamma(t) = \bar h _t$, $k({\bf x-x'}) = 2 D_0 D( {\bf
x-x'})$, $k^{''}(0) = k_{x_i x_i} (0)$ and $k^{''''}(0) = k_{x_i
x_i x_i x_i}$ (0).  Equation  (4) enables us to calculate the
joint PDF $P(\widetilde{h},u_i,p_{ij},t)$ and all the moments $S_n
= \langle\widetilde{h}^n\rangle$ in $d$-dimensions. Various
moments has been calculated explicitly for the three-dimensional
case in Appendix B. To derive a closed expression for $P(h-\bar
h,u_i,t)$ one needs to know the moments such $ < h^n u_i ^m u_j ^l
p_{ij}>$. As appeared in Appendix B, using  eq.(4) one can show
that such moments are identically zero [11]. Using the identity
mentioned above, it can be shown that $P(\widetilde{h},u_i,t)$ has
the following expression ( see appendix C);

 \begin{eqnarray} \label{pp}
P(\widetilde{h},u_i,t)=\int\frac{d\lambda}{2\pi}\prod_i ^d
\frac{d\mu_i}{2\pi}
 \exp(i\lambda(h-\bar{h}(t))+i\sum_{l} ^d\mu_{l}u_{l} )
 \times Z(\lambda,\mu_i,t)
\end{eqnarray}

where,

 \begin{eqnarray} \label{ff}
 Z(\lambda,\mu_i,t) = F_1 (\lambda,\mu_1 )F_2 (\lambda,\mu_2
 )...
 F_d ( \lambda,\mu_d ) \exp{(- \lambda^2 k(0) t)}
\end{eqnarray}

and,

\begin{eqnarray}
 &&F_j( \lambda, \mu_j, {t})= ( 1 -
{\tanh}^{2}(\sqrt{2i{{k}_{{xx}}}(0)\alpha \lambda}{t}))^{-\frac{1}{4}} \nonumber \\
&&\exp[-\frac{i}{2}\alpha k''(0)\lambda t^2
-\frac{1}{2}i\mu_j^{2}\sqrt{\frac{2ik_{xx}(0)}
{\alpha\lambda}}{\tanh}(\sqrt{2i{{k}_{{xx}}}(0)\alpha
\lambda}{t})].\nonumber \\
\end{eqnarray}
\vskip 1cm

Indeed  $Z(\lambda,\mu_i,t)$ is the generating function of the
$\tilde h$ and  $u_i$`s. Therefore expanding eq.(6) in powers of
$\lambda$, all the moments $S_n$ can be derived. For instance,
the first five moments are as follows

\begin{eqnarray}
&&\langle \tilde{h}^{2} \rangle=(\frac{k^2(0)}{\alpha
k''(0)})^\frac{2}{3} [-(\frac{d}{3})(\frac{t}{t^*})^4
+2\frac{t}{t^*}] \cr \nonumber  &&\langle \tilde{h}^{3} \rangle
=-\frac{8d}{15} (\frac{k^2(0)}{\alpha
k''(0)})(\frac{t}{t^*})^{6}\cr \nonumber
&&\langle\tilde{h}^{4}\rangle= (\frac{k^2(0)}{\alpha
k''(0)})^\frac{4}{3} \cr \nonumber\cr &&\hspace{1cm}\times[
(\frac{1}{3} d^2 - \frac{136}{105}d)(\frac{t}{t^*})^8 -4 d
(\frac{t}{t^*})^{5}+12(\frac{t}{t^*})^{2}]\cr \nonumber
&&\langle\tilde{h}^{5}\rangle=- (\frac{k^2(0)}{\alpha
k''(0)})^\frac{5}{3} \cr
\nonumber\\
&&\hspace{1cm}\times[\frac{16 }{9} d ( \frac{248}{105}-d)
(\frac{t}{t^*})^{10}+\frac{32}{3} d(\frac{t}{t^*})^{7}] \nonumber
\end{eqnarray}
where $t_{*}=(\frac{k(0,0)}{\alpha ^2k^{\prime \prime
}{}^2(0,0)})^{1/3}$.

\begin{figure}[htbp]
  \includegraphics[width=0.5\linewidth]{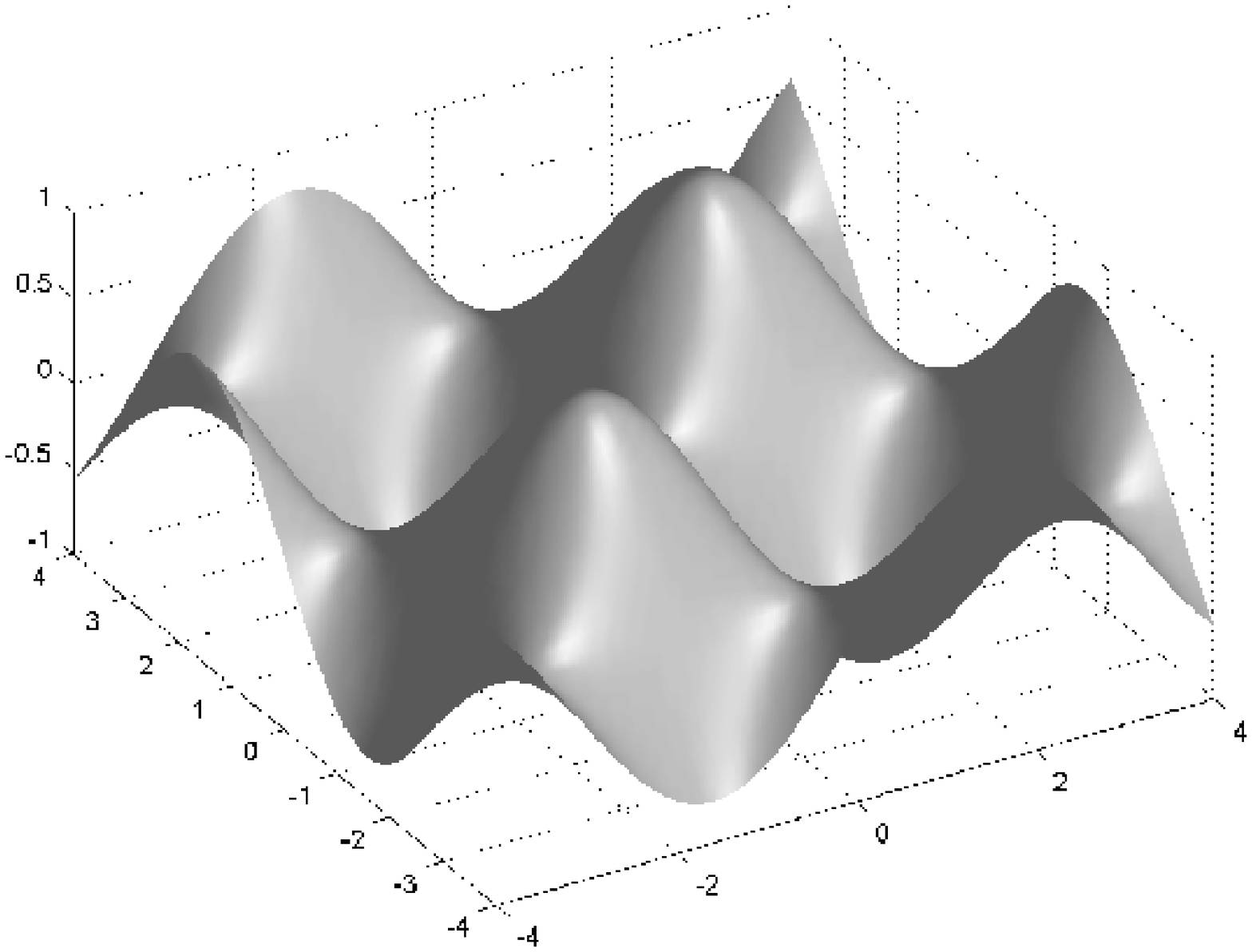}%
  \includegraphics[width=0.5\linewidth]{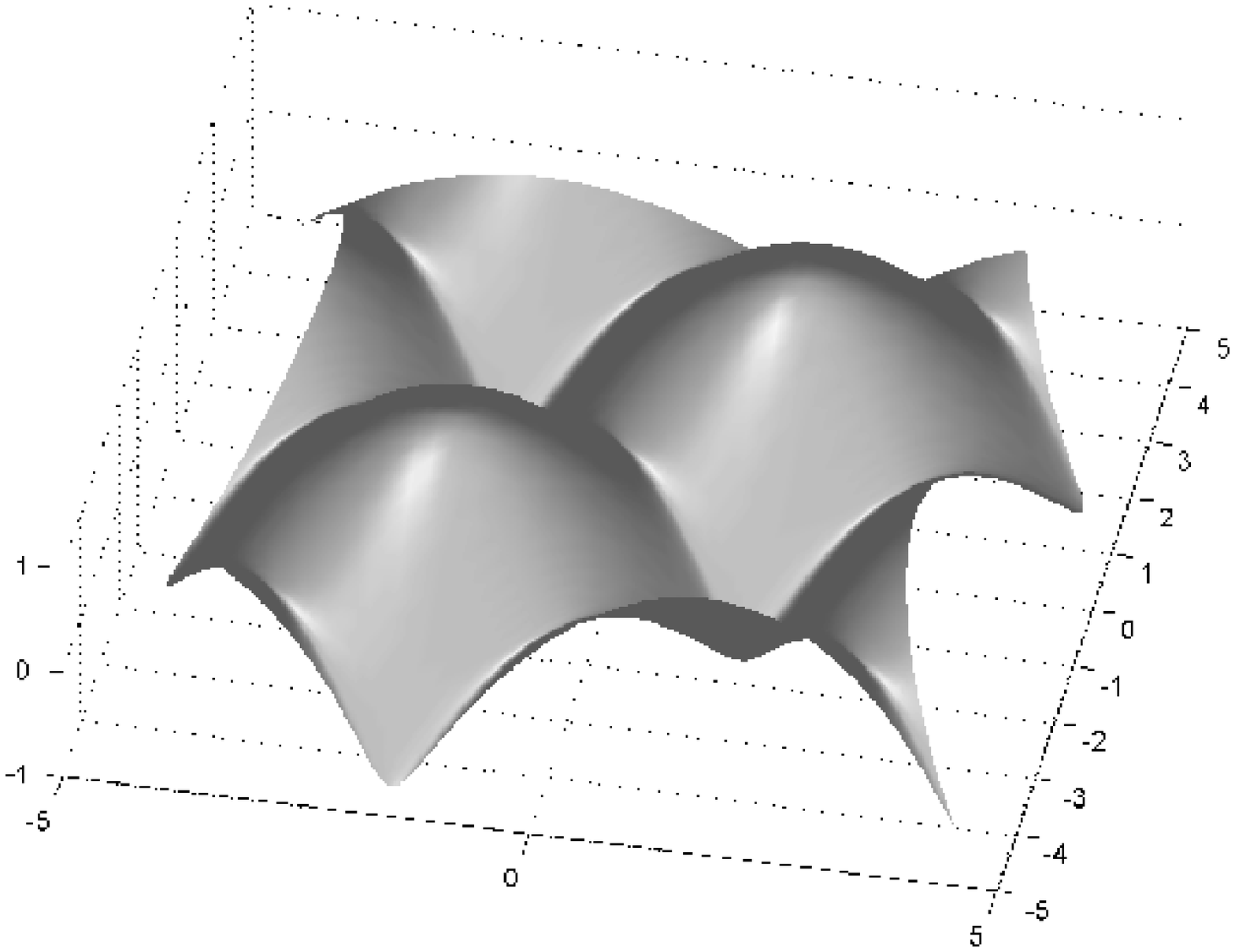}
\includegraphics[width=0.5\linewidth]{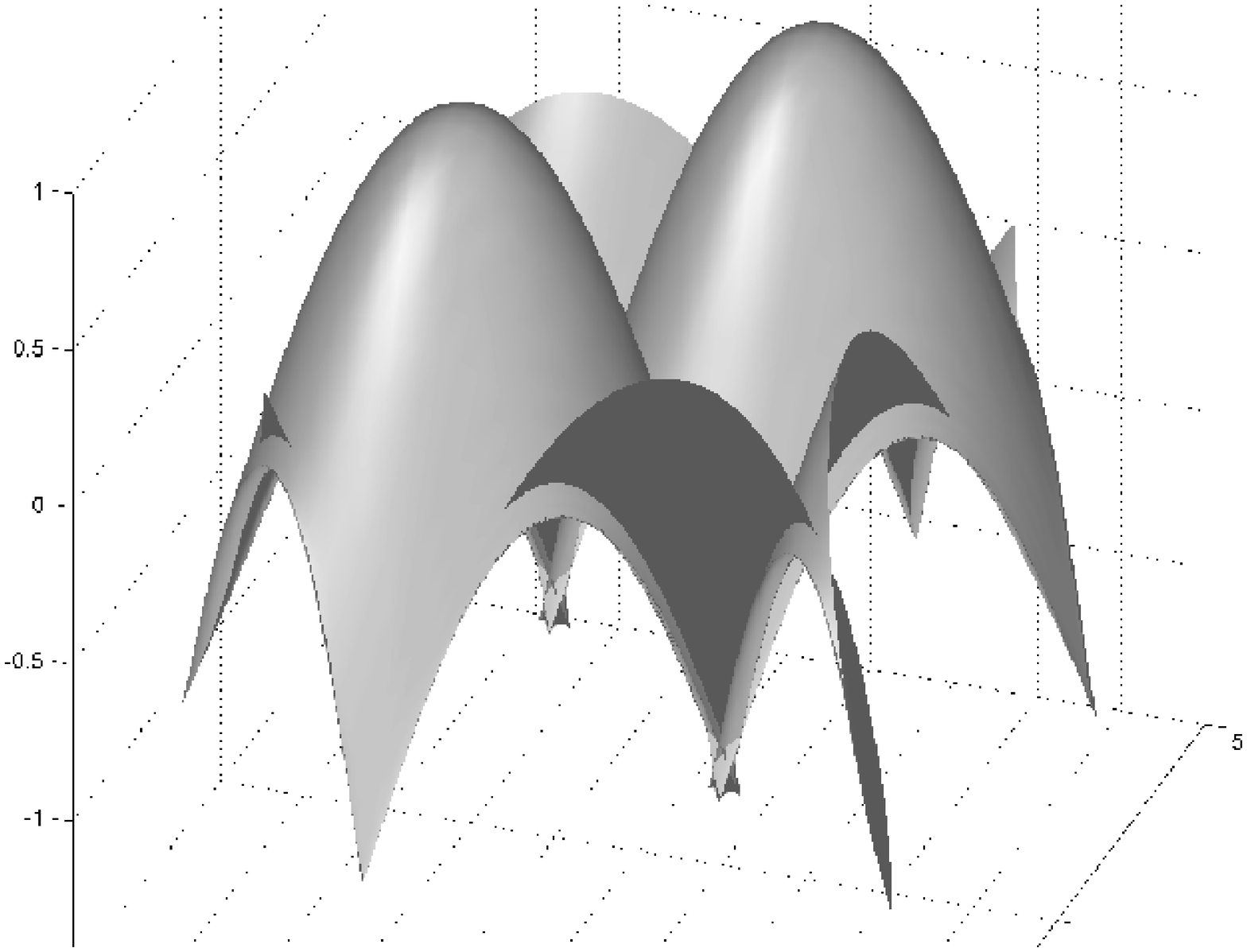}
  \caption[Estimated drift coefficient $D^{(1)}(h_r, r)$]%
{ Different time snapshots of height fields of the zero tension
($\nu =0$), unforced two dimensional KPZ equation with initial
condition $h(x,y,0) = sin(x) sin(y) $. In this case the time scale
$t_{c,\nu=0}$ is equal $1$. In the upper figure the height field
is single valued ( for time scale $t < t_{c,\nu=0}$). In the
middle we demonstrate the height field in time scale
$t=t_{c,\nu=0}$ which is the singularity time scale. In the lower
figure, which is for time scales $t
> t_{c,\nu=0}$, the field $h(x,y)$ is multi-valued.}
  \label{fig:D3}\label{fig:D4}
\end{figure}

The important content of the exact expressions derived above is
that through them the time scale that the height`s PDF
$P(h-{\bar{h}},t)$ lacks positivity condition after $t_c$.
 The positivity of PDF means that all the even moments of $<(h-\bar{h})^{2n}>$ must be positive.
 In fact the above moment
relations indicate that different even order moments become {\it
negative} in some distinct characteristic time scales. Closer
looking at the even moment relations reveals that the higher the
moments are, the smaller their characteristic time scales become
such that asymptotically tends to $t_c= a_d t_*$ for very large
even moments. The coefficients $a_d$ are of order of unity [8,11].
Indeed it can be shown that after time scale $t_c$ the $right$
$tail$ of the probability distribution function (PDF) of height
fluctuations (i.e. $P(h-\bar{h},t)$) is going to become negative,
which is reminiscent of the singularity creation. In what follows
we argue that the two time scales $t_c$ and $t_{c, \nu
=0}$ are related each other and $t_c \simeq t_{c, \nu=0}$ [8,11].

We note that the eq.(6) has the property that $%
Z(0,0,0,t)=1$ which means that $\int_{-\infty }^{+\infty }P(h-\bar{h}%
,u,v;t)d(h-\bar{h})dudv=1$ for {\it every time } $t$ ( $ 0
\preceq t \prec \infty$ ). So the PDF of $h-\bar{h}$ and its
derivatives is always normalizable to unity. In the limit of $\nu
=0$ after $t_{c, \nu=0}$ the height field becomes multi-value on
the valleys, which is related to the left tail of the
$P(h-\bar{h})$. The multiplicity of height field on valleys, on
which the height difference $h-\bar{h}$ is mostly negative,
increases the probability measure in left tail of the PDF.
Therefore to compensate the exceeded measure related to the
multi-valued solutions the right tail of the PDF tails should
become negative. Therefore one concludes that $t_c \simeq t_{c,
\nu=0}$. On the other hand as mentioned in the introduction the
singularities in the limit $\nu \rightarrow 0$ can be constructed
from multi-valued solutions of the KPZ equation with $\nu =0$ by
Maxwell cutting rule [5], which makes the discontinuity in the
derivative of height field. Therefore the time scale that the
zero tension KPZ equation produces  multi-valued solutions is the
same as the time scale of singularity formation in KPZ equation
with infinitesimal surface tension. So $ t_{c,\nu=0} = t_{c,\nu
\rightarrow 0} \simeq t_c$, where $t_c= a_d t_*$ and
$t_{*}=(\frac{k(0,0)}{\alpha ^2k^{\prime \prime
}{}^2(0,0)})^{1/3}$. The $t_c$ scales with $\sigma$ as $
\sigma^{\frac{d+4}{3}}$. Hence the smaller the $\sigma$, the
shorter the time scale of singularity creation.

 Taking into account that $\alpha > 0$ and
$k^{\prime\prime}(0,0) < 0$, the odd order moments $S_{2m+1}$ are
positive in time scales before formation of
multivalued solution. It means that the probability density function $P(h-{\bar h}%
,t)$ in this time regime is positively skewed. Therefore the
probability distribution functions of height difference has a non
zero skewness as it evolves in time, at least up to the time
scale where the multivalued solutions are formed. In Fig.(3-a),
using the equations (4-7), we have numerically sketched the PDF
evolution in time for zero tension $3+1$-dimensional KPZ equation.
As the system evolves in time, the formation of the multivalued
solutions leads to the negativity of the right tail in the PDF.
Also the evolution of PDF right tail  for time scales before and
after $t_c$ can be checked in Fig.(3-b).

Now we would like to add a few comments on the equation governing
the PDF of $h - \bar h$. As shown in eq.(4), for zero tension KPZ
equation, the $P(h, h_x, h_{xx};t)$ satisfies a closed equation.
Adding an infinitesimal surface tension  ( $ \nu \neq 0$) to the
KPZ equation, the PDF equation will no longer  be closed. Here we
have proved that the height`s PDF of zero tension KPZ equation
lacks the positivity condition after the finite time scale $t_c$.
The situation is similar to the results given in [16], where the
derived PDF of velocity increments of {\it invisid} Burgers
equation is not positive definite. In 1+1 dimensions, it has been
already proved that adding an infinitesimal surface tension will
guarantee the positivity of the PDF of $h - \bar h$ [8,17].

The same argument to find the singularity time scale i.e. $t_{c0}$,
can be applied to the problem of decaying tensionless KPZ equation
with random initial condition.  We use the following probability
density functional for initial height field $(h_{0}(x)$ and its
spatial derivative $(u_{0}(x)$ in 1+1 dimension:

\be{\label{ini}} P[h_{0}(x),u_{0}(x)]\propto\exp{\left(-\int dx
dx' h_{0}(x)B(x-x')h_{0}(x') +\int dx dx'
u_{0}(x)B''(x-x')u_{0}(x')\right)}, \ee

where the equality will be hold by a normalization constant and
$B''(x)=B_{xx}(x)$. The initial distribution (8) shows that
initial height field and its derivative are both Zero mean,
statistically homogeneous, Gaussian processes and spatially
correlated with covariance :

\bea
&&\label{i1}<h_{0}(x)h_{0}(x')>=2B(x-x') \\
&&\label{i2}<u_{0}(x)u_{0}(x')>=-2B''(x-x')\\
&&<h_{0}^{n}(x)u_{0}^{m}(x')>=<h_{0}^{n}(x)> <u_{0}^{m}(x')>.
 \eea

We set $B(x)$ to be Gaussian function with standard deviation
$\sigma_{0}$:

\be \label{i3} B({x-x'}) = \frac{1}{\sqrt{(\pi \sigma_{0}^2)}}
\exp(- \frac{ (x-x')^2}{\sigma_{0}^2}), \ee

$\sigma_{0}$ is the correlation length of initial height field.
Now using the KPZ equation and its derivative we can derive the
equations  governing the time evolution of height and height
derivative moments. This procedure has been represented in
Appendix B ( by setting $f=0$ ). Since we are interested in
functional dependence of the time at which the singularities
begin to form, it is sufficient to derive the second moment of
height which can be easily found as:

\be\label{mom} < \tilde{h}^{2} >=-{1\over 2}{\alpha^2}<u_{0}^2>^2
t^2+< \tilde{h_{0}}^{2} >. \ee

Eq.(13) shows that the second height moment would be zero at time
$t^{*}_0=\left({< 2\tilde{h_{0}}^{2} >\over \alpha^2 <
\tilde{u_{0}}^{2} >^2} \right)^{1/2}$. Evaluating $<
\tilde{h_{0}}^{2} >$ and $< \tilde{u_{0}}^{2} >$ from eq.(8) and
using eq.(12) we see that $t^*_0$ scales as $\sigma_{0}^{5/2}$. Due
to the fact that the time scale of singularity formation ($t_{c0} $)
is proportional to $t^*_0$ by a constant of order of unity, the same
conclusion is true for $t_{c0}$. Generalization of the above
argument to higher dimensions is straightforward. It can be easily
shown that $t_{c0} \sim \sigma_{0}^{{d+4}\over2}$. This means that
the smaller their characteristic initial length scales cause the
singularities to be produced at smaller time scales.

\begin{figure}[htbp]
  \includegraphics[width=0.5\linewidth]{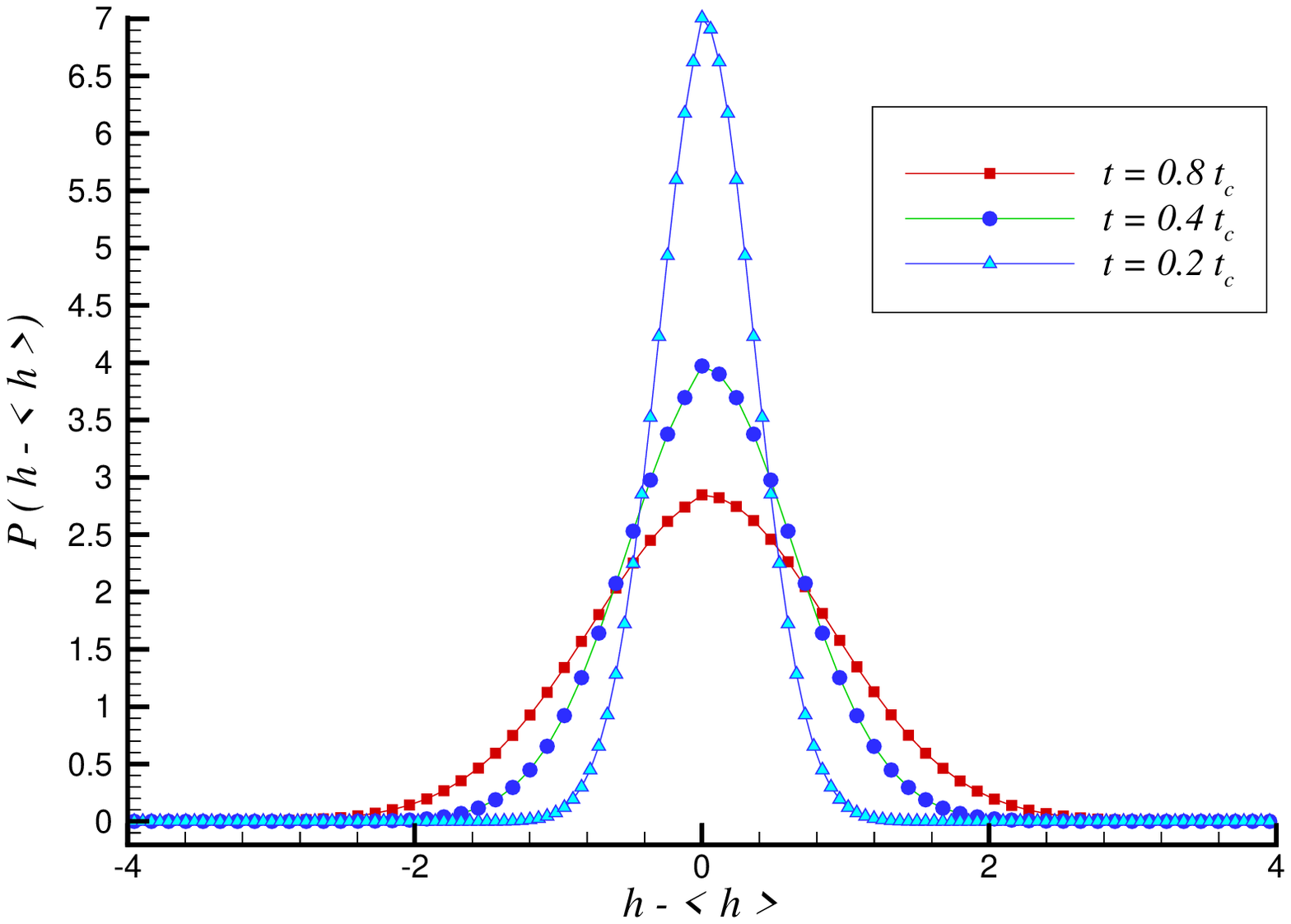}%
  \includegraphics[width=0.5\linewidth]{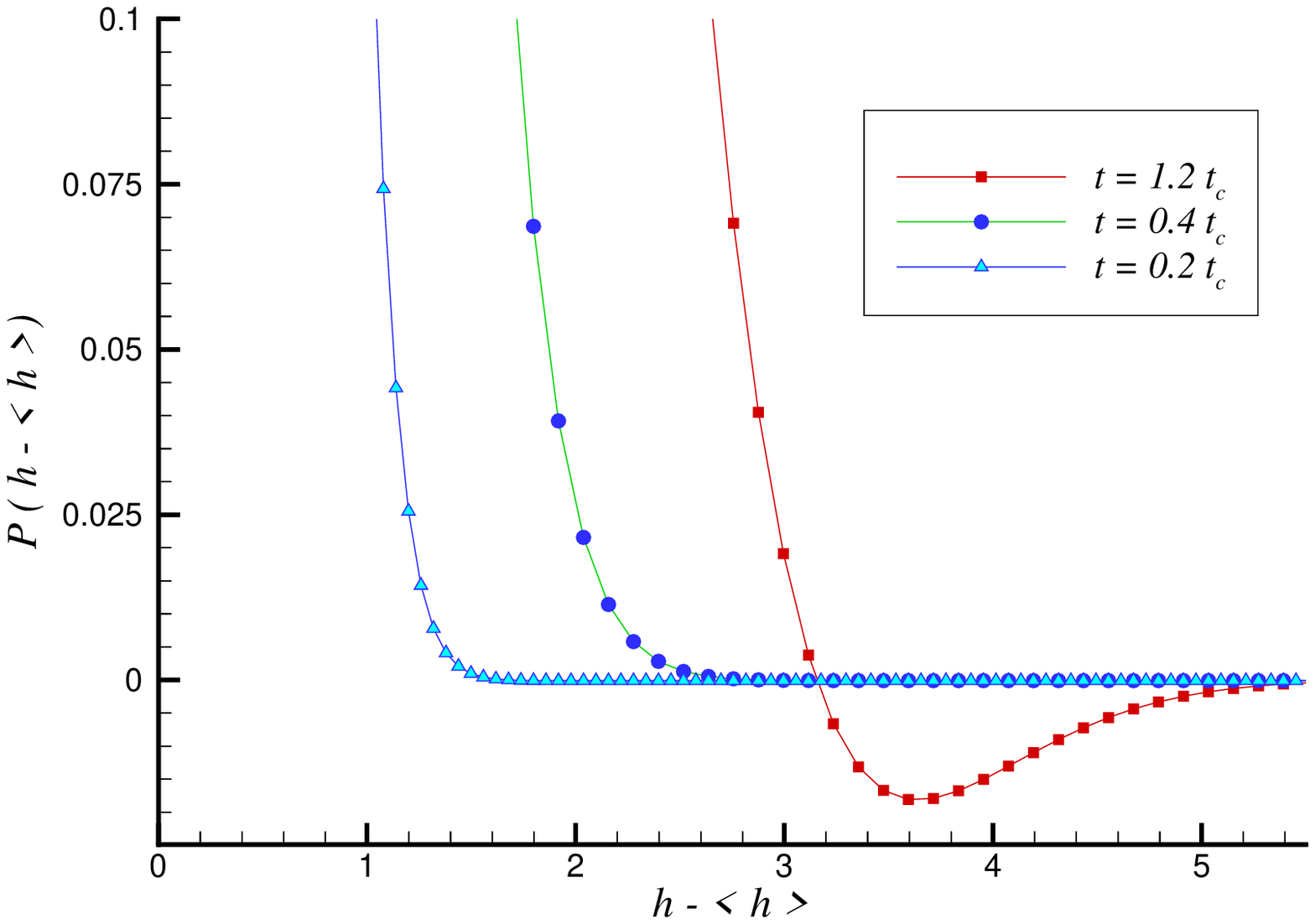}
  \caption[Estimated drift coefficient $D^{(1)}(h_r, r)$]%
  {In the upper graph the time evolution of PDF of $h-\bar h$ before
formation of multivalued solutions at $0.2 t_c$,
 $0.4 t_c$ and $0.8 t_c$ is numerically obtained.
 Lower graph shows the right tails
of the PDF of $h - \bar h$ for $0.2 t_c$, $0.4 t_c$  and $1.2 t_c$
corresponding to time scales before and after formation of
multivalued solutions. }
  \label{fig:D5}\label{fig:D6}
\end{figure}



\newpage

To summarize, we obtain some results on the problem of KPZ
equation in $d+1$ dimensions with a Gaussian forcing, which is
white in time and short-range-correlated in space. In the
non-stationary regime, where the singularities are not yet
developed, we find an exact form for the generating function of
the joint fluctuations of height and height gradients. Starting
from a flat initial condition, we determine the time scale of the
singularity formation and the exact functional form of the time
dependence in the height difference moments at any given order.
We note that if there is no singularity in the height field, the
time evolution of the height's PDF of tensionless KPZ and the KPZ
equation with infinitesimal surface tension are identical. This
is the reason that
 the term $limit _{\nu \rightarrow 0} (\nu \nabla^2 h)$ is equal to
zero for time scales before the singularity time scale $t_c$. We
were able to give the solution of the PDF`s equation for time
scales before $t_c$.

 We believe that the analysis followed in this
paper is also suitable for the zero temperature limit in the
problem of directed polymer in a random potential with short
range correlations [4].
\\

 {\bf Acknowledgments}
\\

We thank J. Davoudi and U. Frisch for useful comments.

\section{APPENDIX A}
In this appendix we prove the equation (4). Define generating
function $Z(\lambda,\mu_i,\eta_{ij},x_i,t)=
<\Theta(\lambda,\mu_{i},\eta_{ij},x_i,t)>$ for the fields $\tilde
h = h - \bar h$, $u_i= h_{x_i}$ and $p_{ij}=h_{x_i x_j}$. The
$\lambda,\mu_{i}$ and $\eta_{ij}$
 are the sources of $\tilde h = h - \bar h$, $u_i= h_{x_i}$ and $p_{ij}=h_{x_i
x_j}$, respectively and $i,j= 1,2, \cdots,d$.

 The explicit expression of $\Theta$ is as follows:
 \begin{eqnarray}
\Theta=
\exp(-i\lambda(h(x,y,z,t)-\bar{h}(t))-i\sum_{i=1}^d\mu_{i}u_{i}-i\sum_{i\leq j=1}^d{\eta_{ij}p_{ij}})\nonumber\\
\end{eqnarray}
where $u_i=h_{x_i}$ and $p_{ij}=h_{x_ix_j}$. Also introduce
$q_{ijk}$ as $q_{ijk}=h_{x_ix_jx_k}$ which will be used later.

By considering the zero-tension KPZ equation we will have the
following time evolution for $h$ and its derivatives;
\begin{eqnarray} \label{kp0}
&&h_t=\frac{\alpha}{2}\sum_{i=1}^3\ u_i^2+f \label{kp0}\\
&&u_{i,t}=\alpha\sum_{l=1}^d u_l p_{li}+f_{x_i}\label{kp2}\\
&&p_{ij,t}=\alpha\sum_{l=1}^d p_{li}p_{lj}+\alpha\sum_{l=1}^du_l
q_{lij}+f_{x_ix_j}. \label{kp3}
\end{eqnarray}

 Also using the Novikov's theorem we can write the following
 identities;
\begin{eqnarray}\label{n1}
&&\langle f\Theta\rangle=-i\lambda
k(0,)Z-i\sum_{l=1}^{3}\eta_{ll}k''(0)Z \label{n1}\\
&&\langle f_{x_i}\Theta\rangle=-i\mu_i
k''(0)Z  \label{n12}\\
&&\langle f_{x_ix_i}\Theta\rangle=-i\lambda
k''(0)Z-i\sum_{l=1}^{3}\eta_{ll}k''''(0)Z \label{n2}\\
&&\langle f_{x_ix_j}\Theta\rangle=-i\eta_{ij}
k''''(0)Z\hspace{1cm}i\neq j \label{n3}
\end{eqnarray}

  where

\begin{eqnarray}
&k(x-x',y-y',z-z')=2D_0D(x-x',y-y',z-z')& ,\nonumber\\
&k(0)=k(0,0,0)=\frac{2D_0}{(\pi \sigma^2)^{\frac{3}{2}}}& ,\nonumber\\
&k_{xxxx}(0,0,0)=\frac{-24D_0}{(\pi \sigma^2)^{\frac{3}{2}}}&  \nonumber\\
&k'(0)=k_x(0,0,0)=k_y(0,0,0)=k_z(0,0,0)=0.\nonumber
\end{eqnarray}

 Differentiate the generating function $Z$ with respect to $t$ and using the eqs.(9-15) and the following identity
\begin{eqnarray}
iZ_{x_l}-i\lambda
Z_{\mu_l}-i\sum_{i}\mu_iZ_{\eta_{il}}\equiv\sum_{i\leq
j}\eta_{ij}\langle q_{ijl}\Theta\rangle ,
\end{eqnarray}

 time evolution of $Z$ can be written simply as

 \begin{eqnarray} \label{zz}
 Z_t&=&i\lambda \gamma(t)Z-i\frac{\lambda\alpha}{2}\sum_l
 Z_{\mu_l\mu_l}-i\alpha\sum_lZ_{\eta_{ll}}
 +i\alpha\sum_{l,i\leq
 j}\eta_{ij}Z_{\eta_{li}\eta_{li}}-\lambda^2 k(0)Z+\sum_l
 \mu_{l}^2k''(0)Z\nonumber\\
 &-&2\lambda\sum_{l}\eta_{ll}k''(0)Z-(\sum_{l,k}\eta_{ll}\eta_{kk}+\sum_{l<k}\eta_{lk}^2)k''''(0)Z
 ,
 \end{eqnarray}
 where $\gamma(t)$ is defined as $\gamma(t)=\overline{h}_t$.
 Fourier transforming  $Z$ respect to $\lambda,\mu_i$ and $
 \eta_{ij}$, the
joint probability density function (PDF)
   of
$\widetilde{h},u_i,p_{ij}$, $P(\widetilde{h},u_i,p_{ij},t)$ is
generated as;
\begin{eqnarray} \label{pp}
 P(\widetilde{h},u_i,p_{ij},t)=\int\frac{d\lambda}{2\pi}\prod_i
 \frac{d\mu_i}{2\pi}\prod_{i\leq j}\frac{d\eta_{ij}}{2\pi}
 \exp(i\lambda(h(x,y,z,t)+\bar{h}(t))+i\sum_{l}\mu_{l}u_{l}+i\sum_{l\leq
 k}{\eta_{lk}p_{lk}})\nonumber\\
 \times Z(\lambda,\mu_i,\eta_{ij},x_i,t)
\end{eqnarray}
 By the use of equations (\ref{zz}) and (\ref{pp}) the equation governing the
 $P(\widetilde{h},u_i,p_{ij},t)$'s time
evolution will be derived as
\begin{eqnarray}  \label{pt}
P_t=\gamma(t)P_{\widetilde{h}}+\frac{\alpha}{2}\sum_l
u_l^2P_{\widetilde{h}}-\alpha (d+2)\sum_lp_{ll}P
-\alpha\sum_{l,k\leq m}p_{lk}p_{lm}P_{p_{km}}+
k(0)P_{\widetilde{h}\widetilde{h}}
-k''(0)\sum_{l}P_{u_lu_l}\hspace{.2cm}\nonumber\\
+2k''(0)\sum_lP_{\widetilde{h}p_{ll}} +k''''(0)\sum_{l\leq
k}P_{p_{lk}p_{lk}} -2k''''(0)\sum_{l<k }P_{p_{ll}p_{kk}}
\end{eqnarray}
where $d$ is the spatial dimension.

 Eq.(\ref{pt}) enables one to write the time evolution of the moments of
  height and it's corresponding derivatives. The obtained equation has the following form

\begin{eqnarray}\label{am}
\frac{d}{dt}\langle\widetilde{h}^{n_0}
AB\rangle=-n_0\gamma(t)\langle\widetilde{h}^{n_0-1}
AB\rangle-\frac{\alpha n_0}{2}\sum_{l}\langle\widetilde{h}^{n_0-1}
ABu_l^2\rangle +\alpha \sum_{l,k\leq
m}n_{km}\langle\widetilde{h}^{n_0}
AB\frac{p_{lk}p_{lm}}{p_{km}}\rangle\nonumber\\
-\alpha\sum_{l}\langle\widetilde{h}^{n_0}
ABp_{ll}\rangle+k(0)n_0(n_0-1)\langle\widetilde{h}^{n_0-2}
AB\rangle
-k''(0)\sum_{l}n_l(n_l-1)\langle\frac{\widetilde{h}^{n_0}AB}{u_l^2}\rangle\nonumber\\
+2k''(0)\sum_{l}n_0n_{ll}\langle\frac{\widetilde{h}^{n_0}AB}{p_{ll}^2}\rangle
+k''''(0)\sum_{l\leq
k}n_{lk}(n_{lk}-1)\langle\frac{\widetilde{h}^{n_0}AB}{p_{lk}^2}\rangle\nonumber\\
+2k''''(0)\sum_{l<k}n_{ll}n_{kk}\langle\frac{\widetilde{h}^{n_0}AB}{p_{ll}p_{kk}}\rangle
\end{eqnarray}

where

\begin{eqnarray}
A=\prod_{i=1}^du_i^{n_i}\nonumber\\
B=\prod_{i\leq j}p_{ij}^{n_{ij}}.\nonumber\end{eqnarray}

By choosing different sort of $n_0$, $n_i$ and $n_{ij}$ values,
various type of coupled differential equations, governing the
evolution
 of the moments, can be constructed .

\section {Appendix B}

In appendix A we have derived the general time evolution equation
for the moments for the arbitrary (d+1)-dimensional case. The
moments in 1+1 and 2+1 dimensions has been derived in refs. [8]
and [11], respectively. Here we are going to restrict ourselves to
the (3+1)-dimensional case. Also the moments in d+1 dimensions is
given at the end of this appendix.

In this appendix the height moments will be calculated exactly and
it will be shown that $\langle h_{x_ix_j}\Theta\rangle,i\neq j$
will be zero by considering flat initial condition in the
(3+1)-dimensional case. However it can be shown that in any
general (d+1)-dimensional case this identity will also be true.
 By choosing different sort of $n_0$, $n_i$ and $n_{ij}$ values,  various
type of coupled differential equations, governing the evolution
 of the moments, can be constructed. The first example is to
choose $n_0=n_i=n_{ij}=0$ then;
\begin{eqnarray}
 -\alpha\sum_l\langle
p_{ll}\rangle=-\alpha\langle \bigtriangledown .\bf
u\rangle=0\Rightarrow\langle \bigtriangledown .\bf u\rangle=0
\end{eqnarray}

 which verifies that the fluid is incompressible in average.
 The main aim is to calculate the moments $\langle p_{ij}\exp(-i\lambda\widetilde{h}-i\sum_l\mu_lu_l)\rangle$.
  Before that, we have to follow several steps. First of all we have to calculate the moments such as
  $\langle p_{ij}u_iu_j\rangle$ and $i \neq j$. Using the eq.(20), we have

 \begin{eqnarray}  \label{can}
\frac{d}{dt}\langle u_iu_jp_{ij}\rangle=\alpha \sum_l\langle
u_iu_jp_{li}p_{lj}\rangle -\alpha \sum_l\langle
u_iu_jp_{ll}p_{ij}\rangle = \alpha \sum_l\langle
u_iu_j(p_{li}p_{lj}-p_{ll}p_{ij}). \rangle
\end{eqnarray}

Looking at the right hand of equation (\ref{can}) we see that the
terms with $l=i$ or $l=j$ cancel each other. Noting that we have
restricted ourselves to the 3+1 dimension, the equation
(\ref{can}) can be written

\begin{eqnarray}  \label{can11}
\frac{d}{dt}\langle u_iu_jp_{ij}\rangle=\langle
u_iu_j(p_{li}p_{lj}-p_{ll}p_{ij}) \rangle \hspace{.5cm}i \neq j
\neq l
\end{eqnarray}

Now inserting the rhs of equation (\ref{can11}) in equation
(\ref{am}), one finds

\begin{eqnarray} \label{a}
&&\frac{d}{dt}\langle u_iu_j(p_{li}p_{lj}-p_{ll}p_{ij}) \rangle=\nonumber\\
&&\alpha\sum_k\langle u_iu_jp_{ki}p_{kl}p_{lj}\rangle
+\alpha\sum_k\langle u_iu_jp_{kj}p_{kl}p_{il}\rangle\nonumber\\
&&-\alpha\sum_k\langle u_iu_jp_{kk}p_{li}p_{lj}\rangle
-\alpha\sum_k\langle u_iu_jp_{kl}p_{kl}p_{ij}\rangle\nonumber\\
&&-\alpha\sum_k\langle u_iu_jp_{ki}p_{kj}p_{ll}\rangle
+\alpha\sum_k\langle u_iu_jp_{kk}p_{ij}p_{ll}\rangle
\end{eqnarray}

It can be easily seen  that in the case that $i\neq j\neq l$ the
rhs of equation (\ref{a}) will be zero. Using this result and
considering a flat initial condition we have;
\begin{eqnarray} \label{a22}
\langle u_iu_j(p_{li}p_{lj}-p_{ll}p_{ij}) \rangle = 0
\end{eqnarray}
and
\begin{eqnarray} \label{a23}
\langle u_iu_jp_{ij}\rangle=0.
\end{eqnarray}

 Also it can be shown by induction that all the moments
such as
$\langle\widetilde{h}^{n_0}p_{ij}^{n_{ij}}u_i^{n_i}u_j^{n_j}\rangle$
are zero too. Therefore one concludes that
 $\langle p_{ij}\exp{(-i\lambda\widetilde{h}-i\sum_l\mu_lu_l)}\rangle = 0 $.
As  shown  in appendix
 C this relation is crucial to derive the eq.(6).

Now let us calculate the moments of  $u_i$`s. Using the eq.(2) it
can be shown that

\begin{eqnarray}\label{a2}
\frac{d}{dt}\langle u_i^n\rangle=-\alpha\langle
u_i^n\sum_lp_{ll}\rangle-n(n-1)k''(0)\langle u_i^{n-2}\rangle.
\end{eqnarray}

 Differentiating $\langle u_i^{n+1}\rangle $ and $\langle
u_i^{n}u_j\rangle $ with respect to $x_i$ and $x_j$ and  using the
statistical homogeneity and equation (26) we have

\begin{equation}\label{a3}
\langle u_i^{n}p_{ii}\rangle=\langle u_i^{n}p_{ii}\rangle=0.
\end{equation}

Therepore;

\begin{eqnarray}\label{a4}
\frac{d}{dt}\langle u_i^{n_i}\rangle=-n_i(n_i-1)k''(0)\langle
u_i^{n_i-2}\rangle
\end{eqnarray}

which is an iterative equation implying that any order of the
$u_i$ moment can be calculated by knowing the lower moments.
Because $\langle u_i\rangle=0$, from equation (\ref{a4}) it is
obvious that any odd moment of $u_i$ will be zero. For $
n_i=2,4,6,8 $, $\langle u_i^{n_i}\rangle$ will be
\begin{eqnarray}\label{a5}
\langle u_i^{2}\rangle=&-2k''(0)t\label{a51}\\
\langle u_i^{4}\rangle=&12k''(0)^2 t^2\label{a52}\\
\langle u_i^{6}\rangle=&-120k''(0)^3 t^3\label{a53}\\
\langle u_i^{8}\rangle=&1680k''(0)^4 t^4\label{a54}.
\end{eqnarray}

and  $\gamma(t)$ ( $\gamma(t)=\overline{h}(t)$) will be;

\begin{eqnarray}\label{a6}
\gamma(t)=\frac{\alpha}{2}\sum_l\langle u_l^2\rangle=-3\alpha
k''(0)t
\end{eqnarray}

For moments such as  $\langle u_i^{n_i}u_j^{n_j}\rangle$, it will
be deduced that
\begin{eqnarray}\label{a6}
\frac{d}{dt}\langle u_i^{n_i}u_j^{n_j}\rangle=
-n_i(n_i-1)k''(0)\langle
u_i^{n_i-2}u_j^{n_j}\rangle\nonumber\\
-n_j(n_j-1)k''(0)\langle u_i^{n_i}u_j^{n_j-2}\rangle-\alpha\langle
u_i^{n_i}u_j^{n_j}\sum_lp_{ll}\rangle
\end{eqnarray}
Differentiate $\langle u_i^{n_i}u_j^{n_j}\rangle$ and $\langle
u_i^{n_i}u_j^{n_j} u_k \rangle$ with respect to  $x_i$  and
$x_k$,respectively one finds
\begin{eqnarray}\label{a7}
\langle u_i^{n_i}u_j^{n_j}\rangle_{x_i}=n_i\langle
u_i^{n_i-1}u_j^{n_j}p_{ii}\rangle+&n_j\langle
u_i^{n_i}u_j^{n_j-1}p_{ij}\rangle  \hskip .5cm i\neq j
\end{eqnarray}
or
\begin{eqnarray}\label{a8}
&\langle u_i^{n_i}u_j^{n_j}u_k\rangle_{x_k}=n_i\langle
u_i^{n_i-1}u_j^{n_j}u_kp_{ik}\rangle+n_j\langle
u_i^{n_i}u_j^{n_j-1}u_kp_{jk}\rangle\nonumber\\
&+\langle u_i^{n_i}u_j^{n_j}p_{kk}\rangle\hspace{.5cm}i\neq j\neq
k.
\end{eqnarray}

Using the statistical homogeneity and identity such as $\langle
u_iu_jp_{ll}\rangle =0$ ( $l \neq i,j$) it can be seen that
equation (\ref{a6}) would be

\begin{eqnarray}\label{a9}
\frac{d}{dt}\langle u_i^{n_i}u_j^{n_j}\rangle=
-n_i(n_i-1)k''(0)\langle
u_i^{n_i-2}u_j^{n_j}\rangle\nonumber\\
-n_j(n_j-1)k''(0)\langle u_i^{n_i}u_j^{n_j-2}\rangle
\end{eqnarray}

Therefore one finds the following expression for the moments;

\begin{eqnarray}
\langle u_i^2u_j^2\rangle=4k''^2(0)t^2\label{a91}\\
\langle u_i^4u_j^2\rangle=-24k''^3(0)t^3\label{a92}\\
\langle u_i^6u_j^2\rangle=240k''^4(0)t^4\label{a94}
\end{eqnarray}

and

\begin{eqnarray}
\langle u_1^2u_2^2u_3^2\rangle=-8k''^3(0)t^3\label{a93}.
\end{eqnarray}

 Now we want to calculate
$\langle\widetilde{h}^2\rangle$ and  its higher moments. Using
eq.(20) it can be shown that the moment $\langle
\widetilde{h}^2\rangle$ satisfies the following equation

\begin{equation}\label{at3}
   \frac{d}{dt}\langle\widetilde{h}^2\rangle=
    -\alpha\sum_{l}\langle\widetilde{h}u^{2}_l\rangle-\alpha\sum_{l}\langle\widetilde{h}^{2}p_{ll}\rangle
    +2k(0)
\end{equation}

But the moment $\langle\widetilde{h}^{2}p_{ll}\rangle$ can be
written as
\begin{equation}\label{p1}
    \langle\widetilde{h}^{2}p_{ll}\rangle=\langle\widetilde{h}^{2}u_{l}\rangle_{x_l}-2\langle\widetilde{h}u^2_{l}\rangle
\end{equation}

where $\langle\widetilde{h}^{2}u_{l}\rangle_{x_l}=0 $, so

\begin{equation}\label{p2}
    \langle\widetilde{h}^{2}p_{ll}\rangle= -2\langle\widetilde{h}u^2_{l}\rangle
\end{equation}

Then equation (\ref{at3}) can be written as

\begin{equation}\label{p3}
    \frac{d}{dt}\langle \widetilde{h}^{2}\rangle=
    \alpha\sum_{l}\langle\widetilde{h}u^{2}_l\rangle
    +2k(0)
    \end{equation}

    Before calculating the $\langle\widetilde{h}u^{2}_l\rangle$
    moment, the more general term
    $\langle\widetilde{h}^{n_0}u_i^{n_i}\rangle$, can be studied, so

    \begin{eqnarray}\label{at2}
    \frac{d}{dt}\langle \widetilde{h}^{n_0}u_i^{n_i}\rangle=
    -n_0 \gamma(t)\langle\widetilde{h}^{n_{0}-1}u_i^{n_i}\rangle
    -\frac{n_0\alpha}{2}\sum_{l}\langle\widetilde{h}^{n_0}u_i^{n_i}u_l^2\rangle\nonumber\\
    -\alpha\sum_{l}\langle\widetilde{h}^{n_0}u_i^{n_i}p_{ll}\rangle
    +n_0(n_0 - 1)k(0)\langle \widetilde{h}^{n_0-2}u_i^{n_i}\rangle\nonumber\\
    -n_i(n_i - 1)k''(0)\langle
    \widetilde{h}^{n_0}u_i^{n_i-2}\rangle\hspace{.5cm}
\end{eqnarray}

where

\begin{eqnarray}\label{at1}
\langle\widetilde{h}^{n_0}u_i^{n_i}p_{ii}\rangle=
-\frac{n_0}{n_i+1}\langle\widetilde{h}^{n_0-1}u_i^{n_i+2}\rangle\\
\langle\widetilde{h}^{n_0}u_i^{n_i}p_{jj}\rangle=
-n_0\langle\widetilde{h}^{n_0-1}u_i^{n_i}u_j^2\rangle\hspace{3mm}i\neq
j
\end{eqnarray}

Therefore equation (\ref{at2}) can be written in a new form as

\begin{eqnarray}\label{qq}
\frac{d}{dt}\langle
\widetilde{h}^{n_0}u_i^{n_i}\rangle=-n_0\gamma(t)\langle
\widetilde{h}^{n_0-1}u_i^{n_i}\rangle+ \frac{\alpha
n_0}{2}\sum_l\langle \widetilde{h}^{n_0-1}u_i^{n_i}u_l^2\rangle\nonumber\\
-\frac{\alpha n_0}{2}\sum_l\delta_{il}\langle
\widetilde{h}^{n_0-1}u_i^{n_i}u_l^2\rangle +n_0(n_-1)k(0)\langle
\widetilde{h}^{n_0-2}u_i^{n_i}\rangle\nonumber\\
-n_i(n_i-1)k''(0)\langle\widetilde{h}^{n_0}u_i^{n_i-2}\rangle\hspace{.5cm}
\end{eqnarray}

From equation (\ref{qq}), $\langle \widetilde{h}u_l^2\rangle$ can
be easily obtained.

\begin{eqnarray}\label{a11}
\frac{d}{dt}\sum_l\langle\widetilde{h}u_l^2\rangle=-\gamma(t)\sum_l\langle
u_l^{2}\rangle\nonumber\\
-\frac{\alpha}{6}\sum_l\langle u_l^4\rangle+\sum_{l<k}\langle
u_l^2u_k^2\rangle
\end{eqnarray}

So finally by using equations (\ref{a51}), (\ref{a52}) and
(\ref{a91}) one gets;

\begin{eqnarray}\label{ah1}
\sum_l\langle \widetilde{h}u_l^2\rangle=-4\alpha k''^2(0)t^3
\end{eqnarray}

Therefore we find;

\begin{eqnarray}\label{a12}
\langle \widetilde{h}^2\rangle=-\alpha ^2k''^2(0)t^4+2k(0)t
\end{eqnarray}

Now it can be shown that the  moment  $
\langle\widetilde{h}^3\rangle$ satisfies the following equation;

\begin{eqnarray}\label{a13}
\frac{d}{dt}\langle\widetilde{h}^3\rangle =
-3\gamma(t)\langle\widetilde{h}^2\rangle-3\frac{\alpha}{2}\sum_l\langle
\widetilde{h}^2u_l^2\rangle -
\alpha\sum_l\langle\widetilde{h}^3p_{ll}\rangle
\end{eqnarray}

 By spatial differentiation
$\langle\widetilde{h}^3p_{ll}\rangle $ will be proportional to $
\langle\widetilde{h}^2u_l^2\rangle$, so

\begin{eqnarray}\label{a14}
\frac{d}{dt}\langle\widetilde{h}^3\rangle
=-3\gamma(t)\langle\widetilde{h}^2\rangle+3\frac{\alpha}{2}\sum_l\langle
\widetilde{h}^2u_l^2\rangle
\end{eqnarray}

and in a similar way the time evolution of  moment  $
\langle\widetilde{h}^4\rangle$
 can be written as
\begin{eqnarray}\label{a15}
\frac{d}{dt}\langle\widetilde{h}^4\rangle
=-4\gamma(t)\langle\widetilde{h}^3\rangle+2\alpha\sum_l\langle
\widetilde{h}^2u_l^2\rangle + 12k(0)\langle\widetilde{h}^2\rangle.
\end{eqnarray}

It is appear that for calculating the moments $
\langle\widetilde{h}^3\rangle$ and $\langle\widetilde{h}^4\rangle
$, we should calculate the  moments $
\langle\widetilde{h}^3u_l^2\rangle$ and $
\langle\widetilde{h}^2u_l^2\rangle$.
   For $ \langle\widetilde{h}^3\rangle$ it is necessary to know $\sum_l
   \langle\widetilde{h}^2u_l^2\rangle$, so

   \begin{eqnarray}\label{a16}
\frac{d}{dt}\sum_l\langle\widetilde{h}^2u^2_l\rangle=
-2\gamma(t)\sum_l\langle\widetilde{h}u^2_l\rangle
-\frac{\alpha}{3}\sum_l\langle\widetilde{h}u^4_l\rangle\nonumber\\
+2\alpha\sum_{l<k}\langle\widetilde{h}u^2_lu^2_k\rangle
+2k(0)\sum_l\langle
u^2_l\rangle-6k''(0)\sum_l\langle\widetilde{h}^2\rangle
\end{eqnarray}

where  $\langle\widetilde{h}u_l^4\rangle$ and $
\langle\widetilde{h}u_l^2u_k^2\rangle$ should be calculated from
the following  equations

\begin{eqnarray}\label{a19}
\frac{d}{dt}\sum_l\langle\widetilde{h}u_l^4\rangle=
-\gamma(t)\sum_l\langle u_l^4\rangle
-\frac{3\alpha}{10}\sum_{l}\langle u_l^6\rangle\nonumber\\
+\frac{\alpha}{2}\sum_{l\neq k}\langle u_l^4u_k^2\rangle
-12k''(0)\sum_{l}\langle\widetilde{h} u_l^2\rangle
\end{eqnarray}

\begin{eqnarray} \label{a17}
\frac{d}{dt}\sum_{l < k}\langle\widetilde{h}u_l^2u_k^2\rangle=
-\gamma(t)\sum_{l < k}\langle u_l^2u_k^2\rangle
-\frac{\alpha}{3}\sum_{l < k}\langle u_l^4u_k^2\rangle\nonumber\\
+\frac{3\alpha}{2} \langle u_1^2u_2^2u_3^2\rangle -
4k''(0)\sum_{l}\langle
\widetilde{h} u_l^2\rangle. \nonumber\\
\end{eqnarray}

 Using the eq.(52) we find;

 \begin{eqnarray}\label{a24}
\langle\widetilde{h}^3\rangle=-\frac{8}{5}\alpha^3k''^3(0)t^6.
\end{eqnarray}

To calculate the moment $\langle\widetilde{h}^4\rangle$ one needs
the following moments

\begin{eqnarray}
&&\langle\widetilde{h}u_1^2u_2^2u_3^2\rangle = -16 \alpha k''^4(0)t^5\label{a25}\\
&&\sum_l\langle\widetilde{h}u_l^6\rangle = -720\alpha k''^4(0)t^5\label{a26}\\
&&\sum_{l<k}\langle u_l^4u_k^4\rangle = 432k''^4t^4\label{a27}\\
&&\sum_l\langle u^2_1u^2_2u^2_3u^2_l\rangle = 144k''^4t^4 \label{a28}\\
&&\sum_{l,k \neq l}\langle u_l^6u_k^2\rangle = 1440k''^4t^4\label{a29}\\
&&\sum_{l,k \neq l}\langle\widetilde{h} u_l^4u_k^2\rangle =
-288k''^4t^5\label{a30}
\end{eqnarray}
\begin{eqnarray}
&&\sum_l\langle\widetilde{h}^2u^4_l\rangle =\frac{364}{5}\alpha^2k''^4(0)t^6 +72k(0)k''^2(0)t^3\label{a31}\\
&&\sum_{l\leq k}\langle \widetilde{h}^2u_l^2u_k^2\rangle = \frac{728}{15}\alpha^2k''^4t^6+ 48 k(0)k''^2(0)t^3\label{a32}\\
&&\sum_{l}\langle \widetilde{h}^3u_l^2\rangle =
\frac{212}{35}\alpha^3k''^4t^7-24 \alpha
k(0)k''^2(0)t^4\label{a33}
\end{eqnarray}
so, finally by substituting the moments above in
equation(\ref{a15}) we find;

\begin{eqnarray}
\langle\widetilde{h}^4\rangle= - \frac{31}{35}\alpha^4k''^4t^8 -
12 \alpha^2 k(0)k''^2(0)t^5 +12 k(0)^2 t^2 \label{a33}
\end{eqnarray}

Generalizing this method to the $d+1$--dimensional case by a
similar amount of calculations, the second,third and fourth
moments can be written as;

 \begin{eqnarray}
&&\langle \tilde{h}^{2} \rangle=(\frac{k^2(0)}{\alpha
k''(0)})^\frac{2}{3} [-(\frac{d}{3})(\frac{t}{t^*})^4
+2\frac{t}{t^*}] \cr \nonumber  &&\langle \tilde{h}^{3} \rangle
=-\frac{8d}{15} (\frac{k^2(0)}{\alpha
k''(0)})(\frac{t}{t^*})^{6}\cr \nonumber
&&\langle\tilde{h}^{4}\rangle= (\frac{k^2(0)}{\alpha
k''(0)})^\frac{4}{3} \cr \nonumber\cr &&\hspace{1cm}\times[
(\frac{1}{3} d^2 - \frac{136}{105}d)(\frac{t}{t^*})^8 -4 d
(\frac{t}{t^*})^{5}+12(\frac{t}{t^*})^{2}]\cr \nonumber
\end{eqnarray}
where $t_{*}=(\frac{k(0,0)}{\alpha ^2k^{\prime \prime
}{}^2(0,0)})^{1/3}$.

\section{Appendix C}

In this appendix using the identities which have found in
appendix B,  joint-probability distribution function (PDF) is
calculated for zero tension KPZ equation. Similar to  appendix B
we restrict ourselves to the 3+1 dimensions case.

The zero tension KPZ equation in 3+1 dimensions has the following
form;

\begin{eqnarray}\label{k1}
 h_t(x,y,z,t)-\frac{\alpha}{2}(h_{x}^2+h_{y}^{2}+h_{z}^{2}) = f
\end{eqnarray}

Now defining

\begin{equation}
h_{x}=u,  h_{y}=v,  h_{z}=w
\end{equation}

 Differentiating the  KPZ equation (\ref{k1}) with respect to
$x$,$y$ and $z$,  we have

\begin{eqnarray}\label{r1}
&h_t=\frac{\alpha}{2}(h_{x}^2+h_{y}^{2}+h_{z}^{2})+f(x,y,t)\\
&u_{t}=\alpha(uu_{x}+vv_{x}+ww_{x})+f_{x}\\
&v_{t}=\alpha(uu_{y}+vv_{y}+ww_{y})+f_{y}\\
&w_{t}=\alpha(uu_{z}+vv_{z}+ww_{z})+f_{z}
\end{eqnarray}

for  h and  corresponding velocity fields. The generating
function $Z(\lambda,\mu_{1},\mu_{2},\mu_{3},x,y,z,t)$ is defined
such as to generate the height and velocity field moments. By
introducing $\Theta$ as
\begin{eqnarray}\label{t1}
\Theta=exp{\left(-i\lambda(h(x,y,z,t)-\bar{h}(t))-i\mu_1u-i\mu_2v
-i\mu_{3}w\right)}
\end{eqnarray}
The generating function will be written as
$Z(\lambda,\mu_{1},\mu_{2},\mu_{3},x,y,z,t)=\langle\Theta\rangle$.

 Using the KPZ equation (\ref{r1}) and its differentiations (78),(79) and (80) with respect to
$x$ , $y$ and $z$, the time evolution of
$Z(\lambda,\mu_{1},\mu_{2},\mu_{3},x,y,z,t)$ can be written as

\begin{eqnarray}\label{evolut}
Z_{t}=i\gamma(t)\lambda
Z-i\lambda\frac{\alpha}{2}\langle(u^2+v^2+w^2)\Theta\rangle
-i\alpha\mu_{1}\langle(uu_{x}+vv_x+ww_x)\Theta\rangle
-i\alpha\mu_{2}\langle(uu_{y}+vv_y+ww_y)\Theta\rangle\nonumber\\
-i\alpha\mu_{3}\langle(uu_z+vv_{z}+ww_z)\Theta\rangle-i\lambda\langle
f\Theta \rangle -i\mu_{1}\langle f_{x}\Theta\rangle
-i\mu_{2}\langle f_{y}\Theta\rangle-i\mu_{3}\langle
f_{z}\Theta\rangle
\end{eqnarray}


where $\gamma(t)=h_t=\frac{\alpha}{2}\langle
u^{2}+v^{2}+w^{2}\rangle$. By considering statistical homogeneity
we have
\begin{eqnarray}\label{h1}
Z_{x}&=&\langle(-i\lambda u-i\mu_{1}u_{x}-i\mu_{2}v_x-i\mu{3}
w_x)\Theta \rangle\\
 Z_{y}&=&\langle(-i\lambda v-i\mu_{1}u_{y}-i\mu_{2}v_{y}-i\mu_{3}w_{y})\Theta \rangle \\
 Z_{z}&=&\langle(-i\lambda w-i\mu_{1} u_{z}-i\mu_{2}v_{z}-i\mu_{3}w_{z})\Theta \rangle
 \end{eqnarray}

 Because we are in a time regime that the singularities has not been formed yet,
 the order of partial derivatives can be exchanged

$\frac{\partial^{2} h}{\partial x_i\partial
x_j}=\frac{\partial^{2} h}{\partial x_j\partial x_i}$ so
$v_x=u_y$,$w_x=u_z$ and $ w_y=v_z$. Keeping the definition of
$\Theta$ (\ref{t1}) in mind, we can easily write
\begin{eqnarray}
i{\frac{\partial}{\partial\mu_{1}}}\langle(-i\mu_{1}
u_{x}-i\mu_{2} v_{x}-i\mu_{3} w_{x})\Theta\rangle=\langle
u_{x}\Theta\rangle-\nonumber \\
i\mu_{1}\langle uu_{x}\Theta\rangle-i\mu_{2}\langle
uv_{x}\Theta\rangle-i\mu_{3}\langle uw_{x}\Theta\rangle.
 \label{h2}
\end{eqnarray}

 From equations (\ref{h1})and (\ref{h2}) we have
\begin{eqnarray}
 \langle u_{x}\Theta\rangle-i\mu_{1}\langle
uu_{x}\Theta\rangle-i\mu_{2}\langle
uv_{x}\Theta\rangle-i\mu_{3}\langle uw_{x}\Theta\rangle
\nonumber\\
=-\lambda\frac{\partial}{\partial\mu_{1}}\langle
u\Theta\rangle=-i\lambda Z _{\mu_{1}\mu_{1}}
 \end{eqnarray}

  And in a similar manner one finds;

\begin{eqnarray}
-i\lambda Z _{\mu_{2}\mu_{2}}=\langle
v_{y}\Theta\rangle-i\mu_{1}\langle
vu_{y}\Theta\rangle-i\mu_{2}\langle
vv_{y}\Theta\rangle-i\mu_{3}\langle vw_{x}\Theta\rangle
\end{eqnarray}

\begin{equation}
-i\lambda Z _{\mu_{3}\mu_{3}}=\langle
w_{z}\Theta\rangle-i\mu_{1}\langle
wu_{z}\Theta\rangle-i\mu_{2}\langle
wv_{z}\Theta\rangle-i\mu_{3}\langle ww_{z}\Theta\rangle
\end{equation}

By using the Novikov's theorem the expression $\langle
f\Theta\rangle$ and $\langle f_{x_i}\Theta\rangle$ can be written
respect to $Z$ ( appeared in  appendix A ). So we have

\begin{eqnarray}
Z_{t}&=&i\gamma(t)\lambda Z-i\lambda\frac{\alpha}{2}
Z_{\mu_{1}\mu_{1}}- i\lambda\frac{\alpha}{2} Z_{\mu_{2}\mu_{2}}-
i\lambda\frac{\alpha}{2} Z_{\mu_{3}\mu_{3}}-\alpha\langle
u_{x}\Theta\rangle-\alpha\langle v_{y}\Theta\rangle-\alpha\langle w_{z}\Theta\rangle\nonumber\\
&-&\lambda^{2}k(0,0,0)Z+\mu_{1}^{2}k_{xx}(0,0,0)Z+\mu_{2}^{2}k_{xx}(0,0,0)Z+\mu_{3}^{2}k_{xx}(0,0,0)Z.
\end{eqnarray}

The $\langle u_{x_i}\Theta\rangle$ terms can be written as

\begin{eqnarray}
\langle
u_{x}\Theta\rangle&=&\frac{i}{\mu_{1}}\langle\Theta\rangle_{x}+\frac{i}{\mu_{1}}\langle
(i\lambda u+i\mu_{2}v_{x}+i\mu_{3}w_{x})\Theta\rangle\nonumber\\
&=&-i\frac{\lambda}{\mu_{1}}Z_{\mu_{1}}-\frac{\mu_{2}}{\mu_{1}}\langle
v_{x}\Theta\rangle-\frac{\mu_{3}}{\mu_{1}}\langle
w_{x}\Theta\rangle,
\end{eqnarray}

and similarly for $\langle v_{y}\Theta\rangle$ and $\langle
w_{z}\Theta\rangle$ we have

\begin{eqnarray}
\langle v_{y}\Theta\rangle=-i\frac{\lambda}{\mu_{2}}Z_{\mu_{2}}
-\frac{\mu_{1}}{\mu_{2}}\langle u_{y}\Theta\rangle
-\frac{\mu_{3}}{\mu_{2}}\langle w_{y}\Theta\rangle
 \end{eqnarray}

 \begin{eqnarray}
 \langle w_{z}\Theta\rangle=-i\frac{\lambda}{\mu_{3}}Z_{\mu_{3}}
-\frac{\mu_{1}}{\mu_{3}}\langle u_{z}\Theta\rangle
-\frac{\mu_{2}}{\mu_{3}}\langle v_{z}\Theta\rangle
 \end{eqnarray}

The terms such as $\langle h_{x_i x_j}\Theta\rangle$ $(i\neq j)$
are the main troublesome terms  which appear in the $Z$`s time
evolution equation ( i.e $\langle u_{y}\Theta\rangle$ ),
preventing us to write the $Z-$equation in a closed form.
Fortunately as it is shown in  appendix B, these terms will
become zero by considering a flat initial condition.

Therefore $Z$ satisfies the following equation;

\begin{eqnarray}
\label{evolut2} Z_{t}=i\gamma(t)\lambda Z-i\lambda\frac{\alpha}{2}
Z_{\mu_{1}\mu_{1}}- i\lambda\frac{\alpha}{2} Z_{\mu_{2}\mu_{2}}-
i\lambda\frac{\alpha}{2} Z_{\mu_{3}\mu_{3}}
+i\alpha\frac{\lambda}{\mu_{1}}Z_{\mu_{1}}-i\alpha\frac{\lambda}{\mu_{2}}Z_{\mu_{2}}\nonumber\\
-i\alpha\frac{\lambda}{\mu_{3}}Z_{\mu_{3}}
-\lambda^{2}k(0,0,0)Z+\mu_{1}^{2}k_{xx}(0,0,0)Z+\mu_{2}^{2}k_{xx}(0,0,0)Z+\mu_{3}^{2}k_{xx}(0,0,0)Z
\end{eqnarray}

In what follows we are going to solve the partial differential
equation above, by using a flat initial condition,
$h(x,y,z,0)=u(x,y,z,0)=v(x,y,z,0)=w(x,y,z,0)=0$, which
equivalently means to write
\begin{eqnarray}
\label{shart} P(\tilde{h},u,v,0)=\delta({\tilde
h})\delta(u)\delta(v)\delta(w).
\end{eqnarray}

This means that;

\begin{eqnarray}
Z(0,0,0,t)=1
\end{eqnarray}

A useful and efficient way to solve the $Z$, time-evolution
differential equation is to factorize it in the following manner
[8,11];

\begin{eqnarray}\label{factor}
&Z(\lambda,\mu_{1},\mu_{2},\mu_{3},t)=F_{1}(\lambda,\mu_{1},t)F_{2}(\lambda,\mu_{2},t)
F_{3}(\lambda,\mu_{3},t)\nonumber\\
&\times\exp{(-\lambda^{2}k(0)t)}
 \end{eqnarray}

Then by inserting (\ref{factor}) in (\ref{evolut2}) we obtain

 \begin{eqnarray}\label{evolut3}
&&{F_{1}}_{t}F_{2}F_{3}+F_{1}{F_{2}}_{t}F_{3}+F_{1}F_{2}{F_{3}}_{t}=\nonumber\\
&&i\gamma(t)\lambda F_{1}F_{2}F_{3} -i\lambda\frac{\alpha}{2}
F_{2}{F_{1}}_{\mu_{1}\mu_{1}}F_{2}F_{3}- i\lambda\frac{\alpha}{2}
F_{1}{F_{2}}_{\mu_{2}\mu_{2}}F_{3}\nonumber\\
&&-i\lambda\frac{\alpha}{2}F_{1}F_{2}{F_{3}}_{\mu_{3}\mu_{3}}
+i\alpha\frac{\lambda}{\mu_{1}}{F_{1}}_{\mu_{1}}F_{2}F_{3}
-i\alpha\frac{\lambda}{\mu_{2}}F_{1}{F_{2}}_{\mu_{2}}F_{3}\nonumber\\
&&-i\alpha\frac{\lambda}{\mu_{3}}F_{1}F_{2}{F_{3}}_{\mu_{3}}
-\lambda^{2}k(0)F_{1}F_{2}F_{2}+\mu_{1}^{2}k''(0,0)F_{1}F_{2}F_{3}\nonumber\\
&&+\mu_{2}k''(0)F_{1}F_{2}F_{2}+\mu_{3}k''(0)F_{1}F_{2}F_{2}.
\end{eqnarray}

Therefore;

\begin{eqnarray}\label{f}
 F_{t}=-i\lambda\frac{\alpha}{2} F_{\mu\mu}+
i\alpha\frac{\lambda}{\mu}F_{\mu}+[\mu^{2}k''(0)-i\alpha\lambda
k''(0)t]F\nonumber\\
\end{eqnarray}

with the initial condition $F(\lambda,\mu,0)=1$.
 This points out
 that the height gradients in the three dimensions evolve
separately from each other before the shock formations, and they
are only coupled with the height field. By Fourier transforming
equation (\ref{f}) respect to $\mu$ a simpler partial differential
equation of order one will appear. This equation can be solved by
the method of Characteristics [8].  Finally the solution of $F$
will be
\begin{eqnarray}
&&F(\mu, \lambda , {t})= ( 1 -
{\tanh}^{2}(\sqrt{2i{{k}_{{xx}}}(0)\alpha \lambda}{t}))^{-\frac{1}{4}} \nonumber \\
&&\exp[-\frac{i}{2}\alpha k''(0)\lambda t^2
-\frac{1}{2}i\mu^{2}\sqrt{\frac{2ik_{xx}(0)}{\alpha\lambda}}{\tanh}(\sqrt{2i{{k}_{{xx}}}(0)\alpha
\lambda}{t})]\nonumber \\
\end{eqnarray}

and

\begin{eqnarray}\label{h}
&&Z(\lambda,\mu_{1},\mu_{2},t)\nonumber \\
&&=F(\lambda,\mu_{1},t)F(\lambda,\mu_{2},t)F(\lambda,\mu_{3},t)\exp{(-\lambda^{2}k(0,0)t)}\nonumber\\
&&=( 1 -{\tanh}^{2}(\sqrt{2i{{k}_{{xx}}}(0)\alpha \lambda}{t}))^{-\frac{3}{4}}\nonumber \\
&&\exp[-\frac{1}{2}i(\mu_1^{2}+\mu_2^{2}+\mu_3^{2})\sqrt{\frac{2ik_{xx}(0)}{\alpha\lambda}}{\tanh}(\sqrt{2i{{k}_{{xx}}}(0)\alpha
\lambda}{t})]\nonumber \\
&&\exp[-\frac{3i}{2}\alpha k''(0)\lambda t^2-k(0)\lambda^2t]
\end{eqnarray}

By inverse Fourier transformation of the generating function $Z$,
the probability distribution function (PDF) of the height
fluctuation can be easily derived;

\begin{eqnarray}
P(\tilde{h},u,v,w,t)= \int\frac{d \lambda}{2\pi}\frac{d
\mu_{1}}{2\pi} \frac{d \mu_{2}}{2\pi}\frac{d \mu_{3}}{2\pi}
\exp(i\lambda\tilde{h}+i\mu_{1}u+i\mu_{2}v+i\mu_{3}w)Z\nonumber\\
\end{eqnarray}

Expanding the solution of the generating function in powers of
$\lambda$, all the $\langle (h-{\bar h})^n \rangle $ moments can
be derived.
 For instance, the first five moments before the sharp
valley formations are
\begin{eqnarray}\label{m1}
&&\langle \tilde{h}^{2}
\rangle=\alpha^2k''^2(0)(\frac{k(0)}{\alpha^2k''^2(0)})^\frac{4}{3}
[-(\frac{t}{t^*})^4 +2\frac{t}{t^*}] \label{m1}\\
&&\langle \tilde{h}^{3} \rangle
=-\frac{8}{5}(\frac{k^{2}(0)}{\alpha^2 k''^2(0)})(\frac{t}{t^*})^{6}\label{m2}\\
&&\langle\tilde{h}^{4}\rangle=\alpha^4k''^4(0)(\frac{k(0)}{\alpha^2k''^2(0)})^\frac{8}{3}\nonumber\\
&&\hspace{1cm}\times[-\frac{31}{35}(\frac{t}{t^*})^8
-12(\frac{t}{t^*})^{5}+12(\frac{t}{t^*})^{2}]\label{m3}\\
&&\langle\tilde{h}^{5}\rangle=-\alpha^5k''^5(0)(\frac{k(0)}{\alpha^2k''^2(0)})^\frac{10}{3}\nonumber\\
&&\hspace{1cm}\times[\frac{-1072}{315}(\frac{t}{t^*})^{10}+32(\frac{t}{t^*})^{7}]\label{m4}
\end{eqnarray}
where $t_* = ( \frac{ k(0,0) } { \alpha^2 k''^2 (0,0) } ) ^{1/3}$.

\end{document}